\documentclass{article}
\usepackage{amsmath}
\usepackage{amssymb}
\usepackage{amsfonts}
\usepackage{amstext}
\usepackage{accents}

\input{tcilatex}
\begin{document}

\author{Cl\'{e}sio E. Mota\thanks{%
clesio200915@hotmail.com} \\
Department of Physics\\
Federal University of Santa Catarina\\
Florian\'{o}polis 88040-900, Santa Catarina,\\
Brazil. \and J. G. Cardoso\thanks{%
jorge.cardoso@udesc.br} \\
Department of Mathematics\\
Centre for Technological Sciences-UDESC\\
Joinville 89223-100, Santa Catarina,\\
Brazil.\\
KEY WORDS: $\varepsilon $-formalism; gravitons; sources; wave equations}
\title{Spin Curvatures in General Relativity and Two-Component Gravitons
with Arbitrary Sources}
\date{ }
\maketitle

\begin{abstract}
We present a short review of the spin curvatures that arise within the
framework of one of the Infeld-van der Waerden formalisms for general
relativity. All the spinor symmetries borne by the inner structure of
general relativity are thus exhibited in an explicit way. The sourceful
extension formulated in an earlier work of the two-component theory of
gravitons in vacuum is concisely reviewed in connection with the
establishment of an important total-symmetry property related to the
energy-momentum conservation law of general relativity.
\end{abstract}

\section{Introduction}

The so-called $\gamma \varepsilon $-formalisms of Infeld and van der Waerden
[1-4] constitute the traditional two-component spinor framework for general
relativity. One of the basic procedures lying behind the construction of
such formalisms involves setting up two pairs of conjugate spin spaces at
every non-singular point of a curved spacetime that is endowed with a
torsionless covariant derivative operator. Remarkably enough, this
construction was carried out much earlier than the establishment of the
definitive conditions for a curved spacetime to admit spinor structures
locally [5,6]. The generalized Weyl gauge group [7] thus operates on any
spin spaces in a way that does not depend at all upon the action of the
relevant spacetime mapping group, and thereby affords the geometric
specifications of any spin objects.

Within the $\gamma \varepsilon $-framework, the crucial assumption
underlying the settlement of any metric and affine structures amounts to
taking at the outset all Hermitian connecting objects, as well as any $%
\varepsilon $-metric spinors, as covariantly constant entities in both the
formalisms. It was shown in Refs. [2,3] that the implementation of this
assumption actually produces in either formalism a self-consistent set of
world-spin metric and affine correlations. The famous Schouten's theory of
spin densities [8,9] turned out, then, to be inherently rooted into the
whole structure of the $\varepsilon $-formalism. The situation concerning
the obtainment of the curvature spinors for the $\varepsilon $-formalism was
slightly touched upon in Refs. [1,10,11], but a fairly complete description
of spin curvatures was given much later in Refs. [2,4]. It thus appeared
that, in either formalism, spin curvatures should arise formally from a
bivector decomposition of mixed world-spin quantities which typically result
out of the local action of covariant derivative commutators on Hermitian
connecting objects.

The Infeld-van der Waerden formalisms were over the years utilized by many
authors not only to reconstruct classical geometric structures, but also to
transcribe classification schemes for world curvature tensors and to treat
Einstein's equations [12-29]. The viability of all such works relies upon
the achievement of systematic developments of the computational methods
involved in the original framework. Amongst these developments, in
particular, is the attainment of a set of algebraic expansions and
valence-reduction devices associated with general spinor-index
configurations, which might be directly applied to any curved spacetimes
because of their intrinsic symbolic character [12]. Notwithstanding the fact
that the formal description of curvature spinors is implicitly carried by
the formalisms, the spin curvatures that occur in some of the geometric
reconstructions we have just mentioned, were obtained in an artificial way
by performing straightforward spinor translations of Riemann and Weyl
tensors. The corresponding translational procedures have notably led to the
earliest spinor version of Einstein's equations [12,16]. However, as in the
case of the popular spin-coefficient Newman-Penrose treatment of Einstein's
equations [12,20], the gauge structure inherently borne by the formalisms
was entirely ruled out and the utmost importance of spin densities
apparently neglected.

Indeed, the most striking physical feature of any curvature spinors lies
over the fact that they are split as sums of purely gravitational and
electromagnetic contributions which bring forth in an inextricably geometric
fashion the occurrence of wave functions for gravitons and photons of both
handednesses. The gravitational contributions for the $\varepsilon $%
-formalism had been employed in Refs. [12,16] to support the transcriptions
performed thereabout, but it had already been established in Ref. [29] that
any of them should be looked upon as a spinor pair which must be associated
to the irreducible decomposition of a Riemann tensor. Gravitational wave
functions for both formalisms are accordingly defined as totally symmetric
curvature pieces that occur in spinor decompositions of Weyl tensors. Any
electromagnetic curvature contribution, on the other hand, amounts to a pair
of suitably contracted pieces\ that enter the spinor representation of a
locally defined Maxwell bivector, while satisfying a peculiar conjugation
property [2,4]. The eventually occurrent covariant constancy of the metric
spinors for the $\gamma $-formalism thus makes up the ultimate condition for
the absence of electromagnetic curvatures from the spacetime geometry
[1,10,15].

The propagation in vacuum of $\varepsilon $-formalism gravitons and their
couplings to \textit{external} electromagnetic fields were described in
Refs. [12,16], whereas the $\gamma \varepsilon $-descriptions of the
couplings between gravitons and Infeld-van der Waerden photons were
exhibited for the first time by the work of Ref. [2] in conjunction with the
derivation of the overall system of sourceless $\gamma \varepsilon $-wave
equations. It was then found that, in both formalisms, any couplings between
gravitons and $\gamma \varepsilon $-photons are strictly borne by the wave
equations that govern the electromagnetic propagation. In Ref. [30], it was
suggested that a description of some of the physical properties of the
cosmic microwave background may be achieved by looking at the propagation in
Friedmann-like conformally flat spacetimes of Infeld-van der Waerden
photons. As carried out by the work of Ref. [2], the explicit derivation of
the wave equations for gravitons was based upon the combination of a
generalized spinor version of the gravitational Bianchi identity with
certain calculational techniques and Einstein's equations in vacuum. All the
statements brought about by the derivation procedures appeared, in effect,
to bear invariance under transformations belonging to the Weyl gauge group,
and the gravitational Bianchi identity effectively supplied the formal field
equations for gravitons. In the $\varepsilon $-formalism, those techniques
take up suitably contracted second-order covariant-derivative operators
which bring forward the geometric specification of the action of
covariant-derivative commutators on arbitrary spin densities.

The introduction of sources into the $\gamma \varepsilon $-descriptions of
gravitons was carried out originally in Ref. [31] where it was shown with
the help of the same calculational techniques as the ones which had been
utilized in the sourceless context, that the second-order covariant
derivative of any energy-momentum tensor held up by the full Einstein's
equations must enjoy a general total-symmetry property related to the
energy-momentum conservation law of general relativity. A work aimed at
describing the effect on the propagation of gravitons of sources coming from
electromagnetic curvatures, was carried out afterwards in Ref. [32]. In this
latter work, all the key electromagnetic patterns that should take place on
the right-hand sides of the wave equations for gravitons in both formalisms,
were clearly obtained.

The leading purpose of the present paper is to review in a concise way the
formulation of the sourceful theory of gravitons as referred to above, while
placing a considerable deal of enhancement on the spinor symmetries tied in
with the inner structure of general relativity. Our work constitutes the
most comprehensive presentation of the traditional two-component spinor
formulation of general relativity, and it is essentially upon this
perspective that the main motivation for having elaborated our review rests.
Nonetheless, we will by this point restrict ourselves to bringing out only
the $\varepsilon $-formalism configurations since this attitude may enable
us to achieve our goals in a more transparent manner.

For the sake of consistency, some basic world-spin definitions are given in
Section 2. In Section 3, we will call upon the curvature structures for the
formalism taken into consideration without recalling any spin-affine
prescriptions. There, we will have to define typical electromagnetic wave
functions but the theory of Infeld-van der Waerden photons [2,4] shall not
be developed. Section 4 replicates the $\varepsilon $-formalism part of the
calculational techniques posed in Ref. [31]. The pertinent field and wave
equations are brought together in Section 5. In Section 6, we make some
remarks on the general relativity framework and set forth the sourceless
limiting case. The inherent spin-density character of the $\varepsilon $%
-formalism, which we had alluded to previously, will be taken for granted to
the extent that we shall specify the gauge behaviours of the $\varepsilon $%
-formalism entities to be allowed for herein without displaying beforehand
Schouten's theory of spin densities.

We should point up that, as far as the development of our work is concerned,
any wave functions must be regarded physically as classical entities. The
notation adhered to in Ref. [2] will be used throughout the paper except
that spacetime components will now be labelled by lower-case Greek letters.
The works of Refs. [1] through [4] shall be utilized many times in what
follows, but we will refer to them explicitly just a few times.

\section{Some basic settings}

We consider a curved spacetime equipped with a symmetric metric tensor $%
g_{\mu \nu }$ of signature $(+---)$ together with a torsionless covariant
derivative operator $\nabla _{\mu }$ that satisfies the metric compatibility
condition%
\begin{equation}
\nabla _{\mu }g_{\lambda \sigma }=0.  \label{1}
\end{equation}%
The Riemann tensor $R_{\mu \nu \lambda \sigma }{}$ of $\nabla _{\mu }$
possesses the general skew symmetries\footnote{%
The whole information on $R_{\mu \nu \lambda \sigma }$ can be particularly
extracted from the world-covariant commutator $[\nabla _{\mu },\nabla _{\nu
}]u^{\lambda }=R_{\mu \nu \sigma }{}^{\lambda }u^{\sigma },$ with $%
u^{\lambda }$ being a world vector.}%
\begin{equation}
R_{\mu \nu \lambda \sigma }{}=R_{[\mu \nu ][\lambda \sigma ]}{},  \label{gr1}
\end{equation}%
and satisfies the Bianchi identity $\nabla _{\lbrack \mu }R_{\nu \lambda
]\sigma \rho }{}=0.$ It also obeys the cyclic identity $R_{[\mu \nu \lambda
]\sigma }{}=0$ such that the index-pair symmetry $R_{\mu \nu \lambda \sigma
}{}=R_{\lambda \sigma \mu \nu }$ likewise holds.\footnote{%
In Ref. [2], this was posed in a somewhat unclear way.} So, the Ricci tensor%
\footnote{%
Our sign convention for $R_{\mu \nu }$ is the same as that adopted in Ref.
[12].} of $\nabla _{\mu }$%
\begin{equation}
R_{\mu \nu }=R_{\mu \tau \nu }{}^{\tau }  \label{2}
\end{equation}%
bears symmetry, whence $R_{\mu \nu \lambda \sigma }$ carries $36-16$ degrees
of freedom while $R_{\mu \nu }$ carries $16-6.$

It is useful to reexpress the cyclic and Bianchi identities as 
\begin{equation}
^{\ast }R^{\lambda }{}_{\mu \nu \lambda }=0,\text{ }\nabla ^{\rho }{}^{\ast
}R_{\rho \mu \lambda \sigma }=0,  \label{11}
\end{equation}%
where%
\begin{equation}
^{\ast }R{}_{\mu \nu \lambda \sigma }=\frac{1}{2}\sqrt{-\mathfrak{g}}%
\mathfrak{e}_{\mu \nu \rho \tau }R^{\rho \tau }{}_{\lambda \sigma }
\label{nLinLin}
\end{equation}%
defines a first-left dual, with $\mathfrak{g}$ denoting the determinant of $%
g_{\mu \nu }$ and $\mathfrak{e}_{\mu \nu \rho \tau }$ standing for one of
the invariant alternating Levi-Civitta world densities. The full Einstein's
equations will be taken as\footnote{%
Equation (\ref{3}) effectively encompasses all the four cases of
presence-absence of sources and presence-absence of the cosmological
constant. It was given originally as Eq. (4.6.32) of Ref. [12].}%
\begin{equation}
\Xi _{\mu \nu }=\frac{1}{2}\kappa (T_{\mu \nu }-\frac{1}{4}Tg_{\mu \nu }),
\label{3}
\end{equation}%
where $T_{\mu \nu }$ amounts to the world version of the energy-momentum
tensor of some sources, $\kappa $ stands for the Einstein gravitational
constant and $T=g^{\lambda \sigma }T_{\lambda \sigma }$ is the metric trace
of $T_{\mu \nu }$. The quantity $(-2)\Xi _{\mu \nu }$ is by definition
identified with the trace-free part of $R_{\mu \nu }$, that is to say,%
\begin{equation}
(-2)\Xi _{\mu \nu }=R_{\mu \nu }-\frac{1}{4}Rg_{\mu \nu },  \label{4}
\end{equation}%
with $R=g^{\lambda \sigma }R_{\lambda \sigma }$ being the Ricci scalar of $%
\nabla _{\mu }.$ In Eq. (\ref{3}), the cosmological constant $\lambda $ is
allowed for implicitly through the equations%
\begin{equation}
G_{\mu \nu }+\lambda g_{\mu \nu }=-\kappa T_{\mu \nu },  \label{add3}
\end{equation}%
where%
\begin{equation}
G_{\mu \nu }=R_{\mu \nu }-\frac{1}{2}Rg_{\mu \nu }  \label{add5}
\end{equation}%
is the Einstein tensor, whence we can make use of the extended trace relation%
\footnote{%
The tensor $G_{\mu \nu }$ equals the trace-reversal part of $R_{\mu \nu }.$
In the absence of the cosmological constant, it obeys the conservation law $%
\nabla ^{\mu }G_{\mu \nu }=0.$}%
\begin{equation}
R=4\lambda +\kappa T.  \label{add10}
\end{equation}

The elements of the Weyl gauge group are non-singular complex $(2\times 2)$%
-matrices whose components are defined as%
\begin{equation}
\Lambda _{A}{}^{B}=\sqrt{\rho }\exp (i\theta )\delta _{A}{}^{B},  \label{5}
\end{equation}%
where $\delta _{A}{}^{B}$ denotes the Kronecker symbol, $\rho $ stands for a
positive-definite differentiable real-valued function of some local
coordinates and $\theta $ amounts to the gauge parameter of the group which
is usually taken as an arbitrary differentiable real-valued function in
spacetime. For the determinant of $(\Lambda _{A}{}^{B})$, we have the
expression 
\begin{equation}
\det (\Lambda _{A}{}^{B})=\Delta _{{\small \Lambda }}=\rho \exp (2i\theta ).
\label{6}
\end{equation}

By definition, any connecting objects for the $\varepsilon $-formalism
satisfy anticommutation relations of the type%
\begin{equation}
\Sigma _{\mu AA^{\prime }}\Sigma _{\nu }^{BA^{\prime }}+\Sigma _{\nu
AA^{\prime }}\Sigma _{\mu }^{BA^{\prime }}=\delta _{A}{}^{B}g_{\mu \nu },
\label{7}
\end{equation}%
which yield the Hermitian relationships%
\begin{equation}
g_{\mu \nu }{}=\Sigma _{\mu }^{AA^{\prime }}\Sigma _{\nu }^{BB^{\prime
}}\varepsilon _{AB}\varepsilon _{A^{\prime }B^{\prime }},\text{ }\varepsilon
_{AB}\varepsilon _{A^{\prime }B^{\prime }}=\Sigma _{AA^{\prime }}^{\mu
}\Sigma _{BB^{\prime }}^{\nu }g_{\mu \nu }{},  \label{8}
\end{equation}%
and thence likewise produce the property\footnote{%
We have sometimes displaced the world index of a connecting object just for
an occasional convenience.}%
\begin{equation}
\Sigma _{\mu A^{\prime }}^{(A}\Sigma _{\nu }^{B)A^{\prime }}=\Sigma
_{A^{\prime }[\mu }^{(A}\Sigma _{\nu ]}^{B)A^{\prime }}=\Sigma _{A^{\prime
}[\mu }^{A}\Sigma _{\nu ]}^{BA^{\prime }}.  \label{9}
\end{equation}%
All Hermitian $\Sigma $-objects enter the metric picture as invariant
spin-tensor densities of absolute weights $\pm 1.$ For instance,%
\begin{equation}
\Sigma _{\mu }^{\prime AA^{\prime }}=\rho \Sigma _{\mu }^{BB^{\prime
}}\Lambda _{B}^{-1}{}^{A}\Lambda _{B^{\prime }}^{-1}{}^{A^{\prime }}=\Sigma
_{\mu }^{AA^{\prime }}.  \label{add500}
\end{equation}

The spinors $(\varepsilon _{AB},\varepsilon _{A^{\prime }B^{\prime }})$ as
well as their inverses $(\varepsilon ^{AB},\varepsilon ^{A^{\prime
}B^{\prime }})$ are skew objects and constitute the only metric spinors for
the $\varepsilon $-formalism. This uniqueness property just stems from the
transformation laws%
\begin{equation}
\varepsilon ^{\prime AB}=\Delta _{{\small \Lambda }}\varepsilon ^{CD}\Lambda
_{C}^{-1}{}^{A}\Lambda _{D}^{-1}{}^{B}=\varepsilon ^{AB},\text{ }\varepsilon
_{AB}^{\prime }=(\Delta _{{\small \Lambda }})^{-1}\Lambda _{A}{}^{C}\Lambda
_{B}{}^{D}\varepsilon _{CD}=\varepsilon _{AB},  \label{10}
\end{equation}%
and their complex conjugates, which tell us that the $\varepsilon $-metric
spinors are invariant spin-tensor densities of weights and antiweights $\pm
1.$ Thus, the spinor structure that represents $\mathfrak{e}_{\mu \nu
\lambda \sigma }$ is spelt out as [12,33]%
\begin{equation}
\sqrt{-\mathfrak{g}}\mathfrak{e}_{AA^{\prime }BB^{\prime }CC^{\prime
}DD^{\prime }}=i(\varepsilon _{AC}\varepsilon _{BD}\varepsilon _{A^{\prime
}D^{\prime }}\varepsilon _{B^{\prime }C^{\prime }}-\text{c.c.})
\label{add12}
\end{equation}%
or, equivalently, as\footnote{%
Any connecting object bears a world-vector character such that the metric
spinors are world invariant quantities as reflected by Eqs. (\ref{8}), (\ref%
{add12}) and (\ref{alt1}).}%
\begin{equation}
\sqrt{-\mathfrak{g}}\mathfrak{e}_{\mu \nu \lambda \sigma }=i(\Sigma _{\mu
AB^{\prime }}\Sigma _{\nu BA^{\prime }}\Sigma _{\lambda }^{AA^{\prime
}}\Sigma _{\sigma }^{BB^{\prime }}-\text{c.c.}),  \label{alt1}
\end{equation}%
with the symbol "c.c." denoting here as elsewhere an overall complex
conjugate piece. The $\varepsilon $-spinors can be seen as Levi-Civitta
densities that obey the relations%
\begin{equation}
\varepsilon _{\lbrack AB}\varepsilon _{C]D}\equiv 0,\text{ }\varepsilon
_{A(B}\varepsilon _{C)D}=\varepsilon _{B(A}\varepsilon _{D)C},  \label{add15}
\end{equation}%
whence we also have%
\begin{equation}
\Delta _{{\small \Lambda }}=\frac{1}{2}\varepsilon ^{AB}\Lambda
_{A}{}^{C}\Lambda _{B}{}^{D}\varepsilon _{CD}.  \label{alt2}
\end{equation}

\section{Curvature spinors}

The information on the overall $\varepsilon $-curvature splitting associated
to $\nabla _{\mu }$ may emerge from%
\begin{equation}
\lbrack \nabla _{\mu },\nabla _{\nu }]\Sigma ^{\lambda DD^{\prime
}}\doteqdot 2\nabla _{\lbrack \mu }(\nabla _{\nu ]}\Sigma ^{\lambda
DD^{\prime }})=0{\large .}  \label{16}
\end{equation}%
Upon expanding the middle configuration of (\ref{16}), we get%
\begin{equation}
(\Sigma ^{\lambda AB^{\prime }}W_{\mu \nu A}{}^{B}+\text{c.c.})+\Sigma
^{\sigma BB^{\prime }}R_{\mu \nu \sigma }{}^{\lambda }=0,  \label{17}
\end{equation}%
where the quantity $W_{\mu \nu AB}{}$ defines an $\varepsilon $-formalism
mixed world-spin object. It follows that, if (\ref{17}) is transvected with $%
\Sigma _{\lambda DB^{\prime }}$, we will obtain 
\begin{equation}
2W_{\mu \nu A}{}^{B}+\delta _{A}{}^{B}W_{\mu \nu A^{\prime }}{}^{A^{\prime
}}=\Sigma _{AB^{\prime }}^{\lambda }\Sigma ^{\sigma BB^{\prime }}R_{\mu \nu
\lambda \sigma }{},  \label{21}
\end{equation}%
whence it can be stated that%
\begin{equation}
4\func{Re}W_{\mu \nu C}{}^{C}=R_{\mu \nu \lambda }{}^{\lambda }\equiv 0,
\label{19}
\end{equation}%
which means that the trace $W_{\mu \nu C}{}^{C}$ amounts to a purely
imaginary quantity.

In Refs. [1,2,4], it becomes manifest that%
\begin{equation}
W_{\mu \nu \lbrack CD]}{}=\frac{1}{2}\varepsilon _{CD}W_{\mu \nu
M}{}^{M}=-iF_{\mu \nu }\varepsilon _{CD},  \label{m1}
\end{equation}%
with $F_{\mu \nu }$ being a Maxwell bivector, whereas the piece%
\begin{equation}
W_{\mu \nu (CD)}{}=\frac{1}{2}\Sigma _{CB^{\prime }}^{\lambda }\Sigma
_{D}^{\sigma B^{\prime }}R_{\mu \nu \lambda \sigma }{}  \label{g1}
\end{equation}%
should supply the gravitational contribution carried by $W_{\mu \nu CD}.$ It
must be stressfully emphasized that Eq. (\ref{9}) really ensures the
genuineness of the spinor symmetry exhibited by (\ref{g1}). One then obtains
the splitting%
\begin{equation}
W_{\mu \nu CD}{}=\frac{1}{2}\Sigma _{CB^{\prime }}^{\lambda }{}\Sigma
_{D}^{\sigma B^{\prime }}R_{\mu \nu \lambda \sigma }{}-iF_{\mu \nu
}\varepsilon _{CD},  \label{15}
\end{equation}%
which thus behaves as an invariant spin-tensor density of weight $-1,$
namely,%
\begin{equation}
W_{\mu \nu CD}^{\prime }{}=(\Delta _{{\small \Lambda }})^{-1}\Lambda
_{C}{}^{L}\Lambda _{D}{}^{M}W_{\mu \nu LM}=W_{\mu \nu CD}.  \label{50}
\end{equation}

Evidently, the index contraction that yields (\ref{19}) from (\ref{21})
brings about annihilation of the gravitational information carried by $%
W_{\mu \nu CD}{}{}$ whilst a symmetrization over the indices $C$ and $D$ of (%
\ref{15}) similarly causes annihilation of the electromagnetic information.
The primary $\varepsilon $-curvature spinors of $\nabla _{\mu }$ arise from
the decomposition of the bivector configuration of (\ref{15}). We have, in
effect, 
\begin{equation}
\Sigma _{AA^{\prime }}^{\mu }\Sigma _{BB^{\prime }}^{\nu }W_{\mu \nu
CD}=\varepsilon _{A^{\prime }B^{\prime }}\omega _{ABCD}+\varepsilon
_{AB}\omega _{A^{\prime }B^{\prime }CD},  \label{51}
\end{equation}%
where 
\begin{equation}
\omega _{ABCD}=\omega _{(AB)CD}\doteqdot \frac{1}{2}\Sigma _{AA^{\prime
}}^{\mu }\Sigma _{B}^{\nu A^{\prime }}W_{\mu \nu CD}  \label{52}
\end{equation}%
and 
\begin{equation}
\omega _{A^{\prime }B^{\prime }CD}=\omega _{(A^{\prime }B^{\prime
})CD}\doteqdot \frac{1}{2}\Sigma _{AA^{\prime }}^{\mu }\Sigma _{B^{\prime
}}^{\nu A}W_{\mu \nu CD}.  \label{53}
\end{equation}%
Due to the gauge character specified by (\ref{50}), the $\omega $-spinors
are subject to the laws%
\begin{equation}
\omega _{ABCD}^{\prime }=(\Delta _{{\small \Lambda }})^{-2}\Lambda
_{A}{}^{L}\Lambda _{B}{}^{M}\Lambda _{C}{}^{R}\Lambda _{D}{}^{S}\omega
_{LMRS}=\omega _{ABCD}  \label{54}
\end{equation}%
and 
\begin{equation}
\omega _{A^{\prime }B^{\prime }CD}^{\prime }=\rho ^{-2}\Lambda _{A^{\prime
}}{}^{L^{\prime }}\Lambda _{B^{\prime }}{}^{M^{\prime }}\Lambda
_{C}{}^{R}\Lambda _{D}{}^{S}\omega _{L^{\prime }M^{\prime }RS}=\omega
_{A^{\prime }B^{\prime }CD}.  \label{55}
\end{equation}

The world gravitational curvatures of $\nabla _{\mu }$ can be completely
reinstated from the pair%
\begin{equation}
\mathbf{G}=(\omega _{AB(CD)},\text{ }\omega _{A^{\prime }B^{\prime }(CD)}).
\label{56}
\end{equation}%
Thus, the elements of the $\mathbf{G}$-pair occur in the spinor expression
for $R_{\mu \nu \lambda \sigma }$ according to the prescription [12,29]%
\begin{equation}
R_{AA^{\prime }BB^{\prime }CC^{\prime }DD^{\prime }}=\hspace{-1pt}%
(\varepsilon _{A^{\prime }B^{\prime }}\varepsilon _{C^{\prime }D^{\prime
}}\omega _{AB(CD)}\hspace{-1pt}+\varepsilon _{AB}\varepsilon _{C^{\prime
}D^{\prime }}\omega _{A^{\prime }B^{\prime }(CD)})+\text{c.c.}.  \label{57}
\end{equation}%
This property was established in Refs. [2,4] by rewriting Eq. (\ref{g1}) as%
\begin{equation}
\frac{1}{2}\Sigma _{CA^{\prime }}^{\lambda }\Sigma _{D}^{\sigma A^{\prime
}}R_{\mu \nu \lambda \sigma }=\Sigma _{M^{\prime }[\mu }^{E}\Sigma _{\nu
]}^{FM^{\prime }}\omega _{EF(CD)}+\Sigma _{M[\mu }^{E^{\prime }}\Sigma _{\nu
]}^{F^{\prime }M}\omega _{E^{\prime }F^{\prime }(CD)},  \label{904}
\end{equation}%
likewise utilizing the metric formulae%
\begin{equation}
\Sigma _{AM^{\prime }}^{\mu }\Sigma _{B}^{\nu M^{\prime }}\Sigma _{\lbrack
\mu }^{EA^{\prime }}\Sigma _{\nu ]A^{\prime }}^{F}=-2\varepsilon
^{(E}{}_{A}\varepsilon ^{F)}{}_{B}  \label{M1}
\end{equation}%
and%
\begin{equation}
\Sigma _{AM^{\prime }}^{\mu }\Sigma _{B}^{\nu M^{\prime }}\Sigma _{\lbrack
\mu }^{ME^{\prime }}\Sigma _{\nu ]M}^{F^{\prime }}=\varepsilon
_{(AB)}\varepsilon ^{E^{\prime }F^{\prime }}\equiv 0,  \label{M2}
\end{equation}%
together with the complex conjugates of (\ref{M1}) and (\ref{M2}). We may
then pick up the individual $\omega $-spinors of (\ref{57}) in agreement
with the coupling schemes%
\begin{equation}
\frac{1}{2}\Sigma _{AM^{\prime }}^{\mu }\Sigma _{B}^{\nu M^{\prime }}\Sigma
_{CA^{\prime }}^{\lambda }\Sigma _{D}^{\sigma A^{\prime }}R_{\mu \nu \lambda
\sigma }{}=-\Sigma _{AM^{\prime }}^{\mu }\Sigma _{B}^{\nu M^{\prime }}\Sigma
_{\lbrack \mu }^{EA^{\prime }}\Sigma _{\nu ]A^{\prime }}^{F}\omega
_{EF(CD)}=2\omega _{AB(CD)}  \label{910}
\end{equation}%
and%
\begin{equation}
\frac{1}{2}\Sigma _{MA^{\prime }}^{\mu }\Sigma _{B^{\prime }}^{\nu M}\Sigma
_{CM^{\prime }}^{\lambda }\Sigma _{D}^{\sigma M^{\prime }}R_{\mu \nu \lambda
\sigma }{}=-\Sigma _{MA^{\prime }}^{\mu }\Sigma _{B^{\prime }}^{\nu M}\Sigma
_{\lbrack \mu }^{AE^{\prime }}\Sigma _{\nu ]A}^{F^{\prime }}\omega
_{E^{\prime }F^{\prime }(CD)}=2\omega _{A^{\prime }B^{\prime }(CD)}.
\label{92}
\end{equation}

The spinor symmetries carried by the entries of the $\mathbf{G}$-pair
correspond to the skew symmetry in the indices of the pairs $\mu \nu $ and $%
\lambda \sigma $ borne by $R_{\mu \nu \lambda \sigma }$. In view of the
index-pair symmetry of $R_{\mu \nu \lambda \sigma },$ one should demand the
additional symmetries%
\begin{equation}
\omega _{AB(CD)}=\omega _{(CD)AB},\text{ }\omega _{A^{\prime }B^{\prime
}(CD)}=\omega _{(CD)A^{\prime }B^{\prime }},  \label{93}
\end{equation}%
whence the only symmetries of any gravitational spinors look like%
\begin{equation}
\omega _{AB(CD)}=\omega _{(AB)(CD)}=\omega _{(CD)(AB)},\text{ }\omega
_{A^{\prime }B^{\prime }(CD)}=\omega _{(A^{\prime }B^{\prime })(CD)}=\omega
_{(CD)(A^{\prime }B^{\prime })}.  \label{94}
\end{equation}%
Therefore, the second entry of the $\mathbf{G}$-pair has to be regarded as a
symmetric trace-free Hermitian entity and, consequently, there is no fixed
prescription for ordering the indices of $\omega _{A^{\prime }B^{\prime
}(CD)}$. We can thus write the tracelessness relation%
\begin{equation}
g_{\mu \nu }\Sigma ^{\mu CA^{\prime }}\Sigma ^{\nu DB^{\prime }}\omega
_{A^{\prime }B^{\prime }(CD)}=0.  \label{s15}
\end{equation}

A cosmological interpretation of the gravitational spinors can be
accomplished by working out Eq. (\ref{94}) and invoking Einstein's
equations. In this connection, it is convenient to reexpress (\ref{57}) as 
\begin{equation}
R_{AA^{\prime }BB^{\prime }CC^{\prime }DD^{\prime }}=(\varepsilon
_{A^{\prime }B^{\prime }}\varepsilon _{C^{\prime }D^{\prime }}\text{X}%
_{ABCD}+\varepsilon _{AB}\varepsilon _{C^{\prime }D^{\prime }}\Xi
_{CA^{\prime }DB^{\prime }})+\text{c.c.},  \label{100}
\end{equation}%
with the X$\Xi $-spinors being defined by 
\begin{equation}
\text{X}_{ABCD}\doteqdot \frac{1}{4}\varepsilon ^{A^{\prime }B^{\prime
}}\varepsilon ^{C^{\prime }D^{\prime }}R_{AA^{\prime }BB^{\prime }CC^{\prime
}DD^{\prime }}=\omega _{AB(CD)}  \label{101}
\end{equation}%
and%
\begin{equation}
\Xi _{CA^{\prime }DB^{\prime }}\doteqdot \frac{1}{4}\varepsilon
^{AB}\varepsilon ^{C^{\prime }D^{\prime }}R_{AA^{\prime }BB^{\prime
}CC^{\prime }DD^{\prime }}=\omega _{A^{\prime }B^{\prime }(CD)}.  \label{102}
\end{equation}%
The symmetries of the X-spinor promptly yield the statement 
\begin{equation}
\varepsilon ^{AD}\text{X}_{A(BC)D}=0\Leftrightarrow \varepsilon ^{BC}\text{X}%
_{B(AD)C}=0,  \label{X1}
\end{equation}%
which, in turn, produces the relations 
\begin{equation}
\varepsilon ^{AD}\text{X}_{ABCD}=\chi \varepsilon _{BC},\text{ }\varepsilon
^{BC}\text{X}_{ABCD}=\chi \varepsilon _{AD},  \label{X2}
\end{equation}%
with%
\begin{equation}
\text{X}_{AB}{}^{AB}=2\chi ,  \label{X3}
\end{equation}%
and $\chi $ standing for a world-spin invariant. By calling for Eqs. (\ref%
{nLinLin}) and (\ref{add12}), we can write down the first-left dual pattern
[12]%
\begin{equation}
^{\ast }R_{AA^{\prime }BB^{\prime }CC^{\prime }DD^{\prime
}}=[(-i)(\varepsilon _{A^{\prime }B^{\prime }}\varepsilon _{C^{\prime
}D^{\prime }}\text{X}_{ABCD}-\varepsilon _{AB}\varepsilon _{C^{\prime
}D^{\prime }}\Xi _{CA^{\prime }DB^{\prime }})]+\text{c.c.},  \label{105}
\end{equation}%
which, by virtue of the first of Eqs. (\ref{11}), leads us to%
\begin{equation}
\varepsilon _{A^{\prime }D^{\prime }}\varepsilon ^{BC}\text{X}%
_{ABCD}=\varepsilon _{AD}\varepsilon ^{B^{\prime }C^{\prime }}\text{X}%
_{A^{\prime }B^{\prime }C^{\prime }D^{\prime }}.  \label{106}
\end{equation}%
Hence, by coupling both sides of (\ref{106}) with $\varepsilon
^{AD}\varepsilon ^{A^{\prime }D^{\prime }}$, one arrives at the equality%
\begin{equation}
\func{Im}\chi =0,  \label{107}
\end{equation}%
while Eqs. (\ref{100}) and (\ref{107}) immediately give $R=8\chi ,$ whence
the X-spinor possesses 
\begin{equation*}
18-8+1=18-6-1=11
\end{equation*}%
real independent components. In addition, from (\ref{100}), we get the
spinor expression for $R_{\mu \nu }$%
\begin{equation}
R_{AA^{\prime }BB^{\prime }}=\frac{1}{4}R\varepsilon _{AB}\varepsilon
_{A^{\prime }B^{\prime }}-2\Xi _{AA^{\prime }BB^{\prime }},  \label{R1}
\end{equation}%
such that, by recalling Eqs. (\ref{3}) and (\ref{4}), we see that the $\Xi $%
-spinor satisfies the full Einstein's equations 
\begin{equation}
\Xi _{AA^{\prime }BB^{\prime }}=\frac{1}{2}\kappa (T_{AA^{\prime }BB^{\prime
}}-\frac{1}{4}T\varepsilon _{AB}\varepsilon _{A^{\prime }B^{\prime }}),
\label{R2}
\end{equation}%
while%
\begin{equation}
\Xi _{\mu \nu }=\Sigma _{\mu }^{AA^{\prime }}\Sigma _{\nu }^{BB^{\prime
}}\Xi _{AA^{\prime }BB^{\prime }}  \label{R3}
\end{equation}%
and $\Xi _{\mu }{}^{\mu }=0,$ in compatibility with (\ref{s15}) and (\ref%
{102}). Then, the $\Xi $-spinor possesses%
\begin{equation*}
18-8-1=4+6-1=10-1=9
\end{equation*}%
real independent components such that the numbers of degrees of freedom of
the Riemann and Ricci tensors are recovered\footnote{%
The prescription $11+9$ was originally given in Ref. [29].} as $11+9$ and $%
1+9,$ respectively. We mention in passing that, when only traceless sources
are present, the spinor expression for the Einstein tensor appears as%
\begin{equation}
G_{AA^{\prime }BB^{\prime }}=-2\Xi _{AA^{\prime }BB^{\prime }}-\lambda
\varepsilon _{AB}\varepsilon _{A^{\prime }B^{\prime }}.  \label{G1}
\end{equation}

The symmetries of X$_{ABCD}$ considerably simplify the four-index reduction
formula [12] 
\begin{align}
\hspace{-0.1cm}\hspace{-0.01cm}\text{X}_{ABCD}\hspace{-0.07cm}& =\hspace{%
-0.07cm}\text{X}_{(ABCD)}-\frac{1}{4}(\varepsilon _{AB}\text{X}%
^{M}{}_{(MCD)}+\varepsilon _{AC}\text{X}^{M}{}_{(MBD)}+\varepsilon _{AD}%
\text{X}^{M}{}_{(MBC)})  \notag \\
& \hspace{-0.07cm}-\frac{1}{3}(\varepsilon _{BC}\text{X}^{M}{}_{A(MD)}+%
\varepsilon _{BD}\text{X}^{M}{}_{A(MC)})-\frac{1}{2}\varepsilon _{CD}\text{X}%
_{AB}{}^{M}{}_{M}.  \label{exp1}
\end{align}%
When combined with (\ref{X1}) and (\ref{X2}), this property provides us with
the expansion%
\begin{equation}
\text{X}_{ABCD}=\text{X}_{(ABCD)}-\frac{1}{12}R\varepsilon _{A(C}\varepsilon
_{D)B},  \label{exp2}
\end{equation}%
which recovers the number of real independent components of X$_{ABCD}$ as $%
10+1,$ with 
\begin{equation}
\text{X}_{(ABCD)}=\text{X}_{A(BCD)}=\text{X}_{(ABC)D}.  \label{add19}
\end{equation}%
In case $T=0,$ we may rewrite (\ref{exp2}) as%
\begin{equation}
\text{X}_{ABCD}=\text{X}_{(ABCD)}-\frac{1}{3}\lambda \varepsilon
_{A(C}\varepsilon _{D)B}.  \label{X5}
\end{equation}%
It is of some interest to notice that the Hermitian configuration 
\begin{align}
& (\varepsilon _{A(C}\varepsilon _{D)B}\varepsilon _{A^{\prime }B^{\prime
}}\varepsilon _{C^{\prime }D^{\prime }})+\text{c.c.}  \notag \\
& =\varepsilon _{AD}\varepsilon _{BC}\varepsilon _{A^{\prime }D^{\prime
}}\varepsilon _{B^{\prime }C^{\prime }}-\varepsilon _{AC}\varepsilon
_{BD}\varepsilon _{A^{\prime }C^{\prime }}\varepsilon _{B^{\prime }D^{\prime
}}  \label{her1}
\end{align}%
gives rise to%
\begin{align}
& \varepsilon _{A^{\prime }B^{\prime }}\varepsilon _{C^{\prime }D^{\prime }}(%
\text{X}_{(ABCD)}-\text{X}_{ABCD})+\text{c.c.}  \notag \\
& =\frac{1}{3}\lambda (\varepsilon _{AD}\varepsilon _{BC}\varepsilon
_{A^{\prime }D^{\prime }}\varepsilon _{B^{\prime }C^{\prime }}-\varepsilon
_{AC}\varepsilon _{BD}\varepsilon _{A^{\prime }C^{\prime }}\varepsilon
_{B^{\prime }D^{\prime }}).  \label{her2}
\end{align}

The electromagnetic spinors carried by $W_{\mu \nu CD}$ comprises the
constituents of the bivector configuration of (\ref{m1}). Formally, we have
the contracted pair%
\begin{equation}
\mathbf{E}=(\omega _{ABC}{}^{C},\text{ }\omega _{A^{\prime }B^{\prime
}C}{}^{C}),  \label{em1}
\end{equation}%
whose entries thus enter the bivector decomposition 
\begin{equation}
\Sigma _{AA^{\prime }}^{\mu }\Sigma _{BB^{\prime }}^{\nu }F_{\mu \nu }=\frac{%
1}{2}i(\varepsilon _{A^{\prime }B^{\prime }}\omega _{ABC}{}^{C}+\varepsilon
_{AB}\omega _{A^{\prime }B^{\prime }C}{}^{C}),  \label{em2}
\end{equation}%
and likewise supply the wave functions%
\begin{equation}
\phi _{AB}=\frac{1}{2}i\omega _{ABC}{}^{C},\text{ }\phi _{A^{\prime
}B^{\prime }}=\frac{1}{2}i\omega _{A^{\prime }B^{\prime }C}{}^{C}.
\label{photon}
\end{equation}%
We are then led to the general curvature splittings 
\begin{equation}
\omega _{ABCD}=\omega _{(AB)(CD)}+\frac{1}{2}\omega
_{(AB)M}{}^{M}\varepsilon _{CD}  \label{em3}
\end{equation}%
and 
\begin{equation}
\omega _{A^{\prime }B^{\prime }CD}=\omega _{(A^{\prime }B^{\prime })(CD)}+%
\frac{1}{2}\omega _{(A^{\prime }B^{\prime })M}{}^{M}\varepsilon _{CD},
\label{em4}
\end{equation}%
along with their complex conjugates. It is shown in Refs. [2,4] that the
spinors of (\ref{em1}) obey the peculiar conjugacy relations%
\begin{equation}
\omega _{ABC}{}^{C}=-\hspace{1pt}\omega _{ABC^{\prime }}{}^{C^{\prime }},%
\text{ }\omega _{A^{\prime }B^{\prime }C}{}^{C}=-\hspace{1pt}\omega
_{A^{\prime }B^{\prime }C^{\prime }}{}^{C^{\prime }}.  \label{em5}
\end{equation}%
From Eqs. (\ref{54}) and (\ref{55}), we see that the electromagnetic
curvatures are invariant spin-tensor densities subject to the laws 
\begin{equation}
\omega _{ABC}^{\prime }{}^{C}=(\Delta _{{\small \Lambda }})^{-1}\Lambda
_{A}{}^{L}\Lambda _{B}{}^{M}\omega _{LMC}{}^{C}=\omega _{ABC}{}^{C}
\label{em6}
\end{equation}%
and 
\begin{equation}
\omega _{A^{\prime }B^{\prime }C}^{\prime }{}^{C}=(\bar{\Delta}_{{\small %
\Lambda }})^{-1}\Lambda _{A^{\prime }}{}^{L^{\prime }}\Lambda _{B^{\prime
}}{}^{M^{\prime }}\omega _{L^{\prime }M^{\prime }C}{}^{C}=\omega _{A^{\prime
}B^{\prime }C}{}^{C}.  \label{em7}
\end{equation}

\section{Calculational techniques}

We shall now present the calculational devices which will indeed pave the
way for the development of Section 5. The starting device is written out as
the commutator%
\begin{equation}
\Sigma _{AA^{\prime }}^{\mu }\Sigma _{BB^{\prime }}^{\nu }[\nabla _{\mu
},\nabla _{\nu }]=\varepsilon _{A^{\prime }B^{\prime }}\Delta
_{AB}+\varepsilon _{AB}\Delta _{A^{\prime }B^{\prime }},  \label{e1}
\end{equation}%
with each of the $\Delta $-kernels amounting to a symmetric second-order
differential operator that bears linearity as well as the Leibniz-rule
property. More explicitly, we have the defining patterns%
\begin{equation}
\Delta _{AB}=\nabla _{C^{\prime }(A}\nabla _{B)}^{C^{\prime }},\text{ }%
\Delta _{A^{\prime }B^{\prime }}=\nabla _{C(A^{\prime }}\nabla _{B^{\prime
})}^{C},  \label{e5}
\end{equation}%
which behave as gauge-invariant spin-tensor densities of weight $-1$ and
antiweight $-1$, respectively. The contravariant form of the $\Delta $%
-operators is supplied by%
\begin{equation}
\Delta ^{AB}=\varepsilon ^{AC}\varepsilon ^{BD}\Delta _{CD},  \label{e6}
\end{equation}%
which immediately produces the configuration\footnote{%
The derivation of (\ref{e7}) includes taking up the covariant constancy of
the $\varepsilon $-spinors.}%
\begin{equation}
\Delta ^{AB}=\nabla _{C^{\prime }}^{(A}\nabla ^{B)C^{\prime }}=-\nabla
^{C^{\prime }(A}\nabla _{C^{\prime }}^{B)}.  \label{e7}
\end{equation}%
When acting on a world-spin scalar, any $\Delta $-operator gives an
identically vanishing result.\footnote{%
In essence, this property reflects the torsionlessness of $\nabla _{\mu }.$}

The object $W_{\mu \nu CD}$ can be alternatively obtained from either of the
commutators%
\begin{equation}
\lbrack \nabla _{\mu },\nabla _{\nu }]\zeta ^{A}=W_{\mu \nu M}{}^{A}\zeta
^{M},\text{ }[\nabla _{\mu },\nabla _{\nu }]\eta _{A}=-W_{\mu \nu
A}{}^{M}\eta _{M},  \label{add101}
\end{equation}%
with $\zeta ^{A}$ and $\eta _{A}$ being some spin vectors. It turns out that
one can write the following derivatives%
\begin{equation}
\Delta _{AB}\zeta ^{C}=\omega _{ABD}{}^{C}\zeta ^{D},\text{ }\Delta
_{A^{\prime }B^{\prime }}\zeta ^{C}=\omega _{A^{\prime }B^{\prime
}D}{}^{C}\zeta ^{D}  \label{e8}
\end{equation}%
and%
\begin{equation}
\Delta _{AB}\eta _{C}=-\hspace{1pt}\omega _{ABC}{}^{D}\eta _{D},\text{ }%
\Delta _{A^{\prime }B^{\prime }}\eta _{C}=-\hspace{1pt}\omega _{A^{\prime
}B^{\prime }C}{}^{D}{}\eta _{D}.  \label{e9}
\end{equation}%
Invoking (\ref{em3}) and (\ref{em4}) then yields the formulae%
\begin{equation}
\Delta _{AB}\zeta ^{C}=\text{X}_{ABD}{}^{C}\zeta ^{D}+\frac{1}{2}\omega
_{ABD}{}^{D}\zeta ^{C}  \label{e10}
\end{equation}%
and%
\begin{equation}
\Delta _{A^{\prime }B^{\prime }}\zeta ^{C}=\Xi _{A^{\prime }B^{\prime
}D}{}^{C}\zeta ^{D}+\frac{1}{2}\omega _{A^{\prime }B^{\prime }D}{}^{D}\zeta
^{C},  \label{e11}
\end{equation}%
together with their complex conjugates and lower-index versions. For a
complex spin-scalar density $\alpha $ of weight $w$, one has the derivatives%
\begin{equation}
\Delta _{AB}\alpha =-w\alpha \omega _{ABC}{}^{C},\text{ }\Delta _{A^{\prime
}B^{\prime }}\alpha =-w\alpha \omega _{A^{\prime }B^{\prime }C}{}^{C},
\label{e16}
\end{equation}%
which come right away from the integrability condition [4]%
\begin{equation}
\lbrack \nabla _{\mu },\nabla _{\nu }]\alpha =2iw\alpha F_{\mu \nu }.
\label{add1}
\end{equation}%
Hence, the prototype for $\Delta $-derivatives of covariant spin-tensor
densities, say, is prescribed as the expansion%
\begin{equation}
\Delta _{AB}(\alpha U_{C...D})=\alpha (\Delta _{AB}U_{C...D}-w\omega
_{ABM}{}^{M}U_{C...D}),  \label{e17}
\end{equation}%
with $U_{C...D}$ being a spin tensor.

There are some situations wherein the calculation of $\Delta $-derivatives
can be performed as if electromagnetic pieces were absent from curvature
splittings such that the differential expansions would then involve only
purely gravitational contributions. The first consideration concerning this
observation is that there would occur a cancellation of those pieces
whenever $\Delta $-derivatives of Hermitian quantities are actually
computed. Such a cancellation would also happen when we let $\Delta $%
-operators act freely upon spin tensors having the same numbers of
contravariant and covariant indices. For $w<0$, it would still occur in the
expansion (\ref{e17}) when $U_{C...D}$ carries $-2w$ indices and $\func{Im}%
\alpha \neq 0$ everywhere. A similar property likewise holds for cases that
involve outer products between contravariant spin tensors and complex
spin-scalar densities having suitable positive weights. We shall see in the
forthcoming Section that such properties neatly fit in with the case of
gravitational wave functions. Also, the derivation of the wave equations to
be shown in the next Section, takes particular account of the differential
splittings%
\begin{equation}
\nabla _{A^{\prime }}^{C}\nabla ^{AA^{\prime }}=\Delta ^{AC}+\frac{1}{2}%
\varepsilon ^{CA}\square ,\text{ }\nabla _{C}^{A^{\prime }}\nabla
_{AA^{\prime }}=\frac{1}{2}\varepsilon _{AC}\square -\Delta _{AC},
\label{e19}
\end{equation}%
along with the simpler algebraic rules 
\begin{equation}
2\nabla _{\lbrack C}^{A^{\prime }}\nabla _{A]A^{\prime }}=\nabla
_{D}^{A^{\prime }}(\varepsilon _{CA}\nabla _{A^{\prime }}^{D})=\varepsilon
_{AC}\square ,\text{ }2\nabla _{A^{\prime }}^{[C}\nabla ^{A]A^{\prime
}}=\nabla _{A^{\prime }}^{D}(\varepsilon ^{AC}\nabla _{D}^{A^{\prime
}})=\varepsilon ^{CA}\square  \label{e20}
\end{equation}%
and the world-spin invariant definition 
\begin{equation}
\square \doteqdot \Sigma _{DD^{\prime }}^{\mu }\Sigma ^{\nu DD^{\prime
}}\nabla _{\mu }\nabla _{\nu }=\nabla _{\mu }\nabla ^{\mu }.  \label{e22}
\end{equation}

\section{Field and wave equations for gravitons}

The totally symmetric piece that occurs in Eq. (\ref{exp2}) defines a
prototype wave function for gravitons, namely,%
\begin{equation}
\Psi _{ABCD}\doteqdot \text{X}_{(ABCD)}.  \label{g2}
\end{equation}%
At each non-singular spacetime point, $\Psi _{ABCD}$ represents the ten
independent degrees of freedom of $g_{\mu \nu }$, and comes into play as a
massless uncharged field of spin $\pm 2$ which is carried by the curvature
structure of a non-conformally flat spacetime. When taken together with its
complex conjugate, the $\Psi $-wave function (\ref{g2}) thus enters the
spinor expression for the Weyl tensor $C_{\mu \nu \lambda \sigma }$ of $%
\nabla _{\mu }$ in accordance with the expansion [12,16]%
\begin{equation}
\Sigma _{AA^{\prime }}^{\mu }\Sigma _{BB^{\prime }}^{\nu }\Sigma
_{CC^{\prime }}^{\lambda }\Sigma _{DD^{\prime }}^{\sigma }C_{\mu \nu \lambda
\sigma }=\varepsilon _{A^{\prime }B^{\prime }}\varepsilon _{C^{\prime
}D^{\prime }}\Psi _{ABCD}+\text{c.c.}.  \label{g3}
\end{equation}

To derive the relevant field equations, one has to utilize the expression (%
\ref{105}) for working out the coupled Bianchi identities%
\begin{equation}
\varepsilon ^{C^{\prime }D^{\prime }}\nabla ^{AA^{\prime }}{}^{\ast
}R_{AA^{\prime }BB^{\prime }CC^{\prime }DD^{\prime }}=0  \label{g4}
\end{equation}%
and 
\begin{equation}
\varepsilon ^{CD}\nabla ^{AA^{\prime }}{}{}^{\ast }R_{AA^{\prime }BB^{\prime
}CC^{\prime }DD^{\prime }}=0.  \label{g5}
\end{equation}%
We have, in effect,%
\begin{equation}
\nabla _{B^{\prime }}^{A}\text{X}_{ABCD}=\nabla _{B}^{A^{\prime }}\Xi
_{CDA^{\prime }B^{\prime }}  \label{g6}
\end{equation}%
whence, performing a symmetrization over the indices $B$, $C$ and $D$ of (%
\ref{g6}) and recalling the property (\ref{add19}), yields the statement 
\begin{equation}
\nabla _{B^{\prime }}^{A}\Psi _{ABCD}=\nabla _{(B}^{A^{\prime }}\Xi
_{CD)A^{\prime }B^{\prime }}.  \label{g8}
\end{equation}%
Hence, on account of Einstein's equations, we obtain%
\begin{equation}
\nabla _{B^{\prime }}^{A}\Psi _{ABCD}=\frac{1}{2}\kappa \nabla
_{(B}^{A^{\prime }}T_{CD)A^{\prime }B^{\prime }}.  \label{g9}
\end{equation}%
We notice that the symmetrization occurring in (\ref{g8}) annihilates the $T$%
-trace piece of Einstein's equations (see Eq. (\ref{R2}) above). Equation (%
\ref{g9}) comes about as a sourceful field equation for gravitons. In a
similar way, for $\Psi ^{ABCD}$, we get\footnote{%
Equation (\ref{54}) shows us that $\Psi _{ABCD}$ and $\Psi ^{ABCD}$ are
invariant spin-tensor densities of weights $-2$ and $+2$, respectively.}%
\begin{equation}
\nabla _{AB^{\prime }}\Psi ^{ABCD}=-\frac{1}{2}\kappa \nabla ^{A^{\prime
}(B}T^{CD)}{}_{A^{\prime }B^{\prime }}.  \label{e34}
\end{equation}

The gauge behaviour specified by Eq. (\ref{54}) allows us to say that the
valence pattern of $\Psi _{AB}{}^{CD}$ yields an invariant \textit{%
spin-tensor} character. Consequently, if we account for the symmetry of $%
T_{\mu \nu }$ along with the expansion\footnote{%
The spinor version of a traceless $T_{\mu \nu }$ enjoys the same absence of
index ordering prescription as that of $\omega _{A^{\prime }B^{\prime
}(CD)}. $}%
\begin{equation}
T_{CDA^{\prime }B^{\prime }}=T_{(CD)(A^{\prime }B^{\prime })}+\frac{1}{4}%
T\varepsilon _{CD}\varepsilon _{A^{\prime }B^{\prime }},  \label{e36}
\end{equation}%
and make use of the properties%
\begin{equation}
T_{(CD)A^{\prime }B^{\prime }}=T_{(CD)(A^{\prime }B^{\prime })},\text{ }%
T_{CD(A^{\prime }B^{\prime })}=T_{(CD)(A^{\prime }B^{\prime })},  \label{e37}
\end{equation}%
we will obtain the field equation%
\begin{equation}
\nabla _{B^{\prime }}^{A}\Psi _{AB}{}^{CD}=\frac{1}{6}\kappa (\nabla
_{B}^{A^{\prime }}T^{(CD)}{}_{(A^{\prime }B^{\prime })}+2\nabla ^{A^{\prime
}(C}T_{B}{}^{D)}{}_{(A^{\prime }B^{\prime })}).  \label{150}
\end{equation}

In fact, towards carrying through the derivation of the wave equation for
any $\Psi $-field, one must necessarily implement the techniques considered
in Section 4. We shall work out only the structures for the field of Eq. (%
\ref{150}). The statements for the fields carried by Eqs. (\ref{g9}) and (%
\ref{e34}) may be attained thereafter either by making use of the covariant
constancy of the $\varepsilon $-metric spinors or by utilizing the
weight-valence properties of $\Delta $-derivatives mentioned in Section 4.
In this Section, we will use the particular convention whereby vertical bars
surrounding an index block should mean that the indices singled out are not
to partake of a symmetry operation.

Let us rewrite the kernel of the right-hand side of Eq. (\ref{150}) as%
\begin{equation}
\varepsilon ^{CM}\varepsilon ^{DN}\varepsilon ^{B^{\prime }D^{\prime
}}\nabla _{(B}^{A^{\prime }}T_{MN)A^{\prime }D^{\prime }}=-\varepsilon
^{CM}\varepsilon ^{DN}\varepsilon ^{D^{\prime }B^{\prime }}\varepsilon
_{S(B}\nabla ^{SA^{\prime }}T_{MN)A^{\prime }D^{\prime }}.  \label{e39}
\end{equation}%
By operating on (\ref{e39}) with $\nabla _{B^{\prime }}^{L}$ and using the
index-displacement rule%
\begin{equation}
\nabla ^{L(A^{\prime }}\nabla ^{D^{\prime })S}=\nabla ^{S(A^{\prime }}\nabla
^{D^{\prime })L}+\varepsilon ^{LS}\Delta ^{A^{\prime }D^{\prime }},
\label{e40}
\end{equation}%
one gets the development%
\begin{eqnarray}
&&\varepsilon ^{CM}\varepsilon ^{DN}\varepsilon _{S(B}\nabla ^{LD^{\prime
}}\nabla ^{SA^{\prime }}T_{MN)A^{\prime }D^{\prime }}  \notag \\
&=&\varepsilon ^{CM}\varepsilon ^{DN}\varepsilon _{S(B}\nabla ^{L(A^{\prime
}}\nabla ^{D^{\prime })S}T_{MN)A^{\prime }D^{\prime }}  \notag \\
&=&\varepsilon ^{CM}\varepsilon ^{DN}\varepsilon _{S(B}(\nabla ^{S(A^{\prime
}}\nabla ^{D^{\prime })L}+\varepsilon ^{LS}\Delta ^{A^{\prime }D^{\prime
}})T_{MN)A^{\prime }D^{\prime }}  \notag \\
&=&\varepsilon ^{CM}\varepsilon ^{DN}(\nabla _{(B}^{(A^{\prime }}\nabla
^{D^{\prime })L}+\varepsilon ^{L}{}_{(B}\Delta ^{A^{\prime }D^{\prime
}})T_{MN)A^{\prime }D^{\prime }}.  \label{e41}
\end{eqnarray}%
Owing to Eqs. (\ref{X1}) and (\ref{e37}), each of the X-terms taken up by
the evaluation of the $\Delta $-derivative of (\ref{e41}) is equal to zero,
whereas the involved contracted $\omega $-pieces cancel out one another
because of the Hermiticity of $T_{AA^{\prime }BB^{\prime }}.$ We then end up
with%
\begin{equation}
\Delta ^{A^{\prime }D^{\prime }}T_{MNA^{\prime }D^{\prime }}=-2\Xi
^{AA^{\prime }D^{\prime }}{}_{(M}{}T_{N)AA^{\prime }D^{\prime }}=-\kappa
T^{AA^{\prime }D^{\prime }}{}_{(M}{}T_{N)AA^{\prime }D^{\prime }}\equiv 0,
\label{e42}
\end{equation}%
along with the contribution%
\begin{equation}
\frac{1}{2}\kappa \varepsilon ^{CM}\varepsilon ^{DN}\varepsilon _{S(B}\nabla
^{LD^{\prime }}\nabla ^{SA^{\prime }}T_{MN)A^{\prime }D^{\prime }}=\frac{1}{2%
}\kappa \varepsilon ^{CM}\varepsilon ^{DN}\nabla _{(B}^{(A^{\prime }}\nabla
^{D^{\prime })L}T_{MN)A^{\prime }D^{\prime }}.  \label{91}
\end{equation}

For the left-hand side of Eq. (\ref{150}), we utilize the splittings (\ref%
{e19}) to obtain%
\begin{equation}
\nabla _{B^{\prime }}^{L}\nabla ^{AB^{\prime }}\Psi _{AB}{}^{CD}=(\Delta
^{AL}+\frac{1}{2}\varepsilon ^{LA}\square )\Psi _{AB}{}^{CD}.  \label{e44}
\end{equation}%
The $\Delta $-derivative of (\ref{e44}) shows up as [4,31]%
\begin{align}
\Delta ^{AL}\Psi _{AB}{}^{CD}& =(\text{X}^{AL}{}_{M}{}^{C}\Psi _{AB}{}^{MD}+%
\text{X}^{AL}{}_{M}{}^{D}\Psi _{AB}{}^{CM})  \notag \\
& -(\text{X}^{AL}{}_{A}{}^{M}\Psi _{MB}{}^{CD}+\text{X}^{AL}{}_{B}{}^{M}\Psi
_{AM}{}^{CD})  \notag \\
& =\frac{1}{4}R\Psi {}^{CDL}{}_{B}-3Q^{(CDL)M}{}\varepsilon _{MB}  \notag \\
& =\frac{1}{4}R\Psi {}^{CDL}{}_{B}-3Q^{(CDLM)}{}\varepsilon _{MB},
\label{e61}
\end{align}%
where%
\begin{equation}
Q^{CDLM}{}\doteqdot \Psi _{GH}{}^{CD}\Psi {}^{LMGH}{}.  \label{e62}
\end{equation}%
Therefore, by combining together Eqs. (\ref{91})-(\ref{e62}), after lowering
the index $L$ and making some index rearrangements, we get the equation%
\footnote{%
A mistyped sign occurs on the right-hand side of Eq. (44) of Ref. [31] but
it was not passed on to the manipulations that led to any key equations. The
factor $-\frac{1}{2}\kappa $ borne therein by the right-hand sides of Eqs.
(48), (50) and (51) should be replaced with $-\kappa $.}%
\begin{equation}
(\square +\frac{1}{2}R)\Psi {}_{AB}{}^{CD}-6\Psi _{MN}{}^{(CD}\Psi
{}^{GH)MN}\varepsilon _{GA}\varepsilon _{HB}=-\kappa S_{AB}{}^{CD},
\label{e47}
\end{equation}%
with the definition%
\begin{equation}
S_{ABCD}=\nabla _{(B}^{(A^{\prime }}\nabla _{\mid A\mid }^{B^{\prime
})}T_{CD)A^{\prime }B^{\prime }}.  \label{e99}
\end{equation}

It is evident that the configuration (\ref{e99}) bears symmetry in the
indices $B,$ $C$ and $D$. Its symmetry in $A$ and $B$, as presumably imposed
by Eq. (\ref{e47}), can be easily verified by recalling (\ref{e37}) and (\ref%
{e42}) to carry out the computation%
\begin{eqnarray*}
&&\varepsilon _{L[A}(\nabla _{B]}^{(A^{\prime }}\nabla ^{B^{\prime
})L}T_{CDA^{\prime }B^{\prime }}+\nabla _{\mid C\mid }^{(A^{\prime }}\nabla
^{B^{\prime })L}T_{B]DA^{\prime }B^{\prime }}+\nabla _{\mid D\mid
}^{(A^{\prime }}\nabla ^{B^{\prime })L}T_{B]CA^{\prime }B^{\prime }}) \\
&=&\frac{1}{2}\varepsilon _{BA}(\Delta ^{A^{\prime }B^{\prime
}}T_{CDA^{\prime }B^{\prime }}+\nabla _{C}^{(A^{\prime }}\nabla ^{B^{\prime
})L}T_{LDA^{\prime }B^{\prime }}+\nabla _{D}^{(A^{\prime }}\nabla
^{B^{\prime })L}T_{LCA^{\prime }B^{\prime }}) \\
&=&\frac{1}{2}\varepsilon _{BA}(\nabla _{C}^{A^{\prime }}\nabla ^{LB^{\prime
}}T_{LB^{\prime }DA^{\prime }}+\nabla _{D}^{A^{\prime }}\nabla ^{LB^{\prime
}}T_{LB^{\prime }CA^{\prime }})=\varepsilon _{BA}\nabla _{(C}^{A^{\prime
}}\nabla ^{\mu }T_{D)A^{\prime }\mu }=0,
\end{eqnarray*}%
with the last step of which being due to the presupposed divergencelessness
of $T_{\mu \nu }.$ Moreover, this latter symmetry property remains also
valid for the indices $A$, $C$ and $A$, $D.$ We have, for instance,%
\begin{eqnarray*}
&&\varepsilon _{L[A}(\nabla _{\mid B\mid }^{(A^{\prime }}\nabla ^{B^{\prime
})L}T_{C]DA^{\prime }B^{\prime }}+\nabla _{C]}^{(A^{\prime }}\nabla
^{B^{\prime })L}T_{BDA^{\prime }B^{\prime }}+\nabla _{\mid D}^{(A^{\prime
}}\nabla ^{B^{\prime })L}T_{B\mid C]A^{\prime }B^{\prime }}) \\
&=&\frac{1}{2}\varepsilon _{CA}(\nabla _{B}^{(A^{\prime }}\nabla ^{B^{\prime
})L}T_{LDA^{\prime }B^{\prime }}+\Delta ^{A^{\prime }B^{\prime
}}T_{BDA^{\prime }B^{\prime }}+\nabla _{D}^{(A^{\prime }}\nabla ^{B^{\prime
})L}T_{BLA^{\prime }B^{\prime }}) \\
&=&\frac{1}{2}\varepsilon _{CA}(\nabla _{B}^{A^{\prime }}\nabla ^{LB^{\prime
}}T_{LB^{\prime }DA^{\prime }}+\nabla _{D}^{A^{\prime }}\nabla ^{LB^{\prime
}}T_{LB^{\prime }BA^{\prime }})=\varepsilon _{CA}\nabla _{(B}^{A^{\prime
}}\nabla ^{\mu }T_{D)A^{\prime }\mu }=0,
\end{eqnarray*}%
whence $S_{ABCD}$ bears the total-symmetry property%
\begin{equation}
S_{ABCD}=\nabla _{(A}^{(A^{\prime }}\nabla _{B}^{B^{\prime
})}T_{CD)A^{\prime }B^{\prime }}=S_{(ABCD)}.  \label{s3}
\end{equation}

As was said before, $\Psi _{ABCD}$ and $\Psi ^{ABCD}$ are invariant
spin-tensor densities of weights $-2$ and $+2,$ whence they match some of
the situations explained in Section 4 regarding eventual absences of
electromagnetic contributions from $\Delta $-derivatives. It follows that
the left-hand sides of Eqs. (\ref{g9}) and (\ref{e34}) could be treated in a
way quite similar to the case of (\ref{150}). We thus obtain%
\begin{equation}
(\square +\frac{1}{2}R)\Psi {}_{ABCD}{}-6\Psi _{MN(AB}{}\Psi
{}_{CD)}{}^{MN}=-\kappa \nabla _{(A}^{(A^{\prime }}\nabla _{B}^{B^{\prime
})}T_{CD)A^{\prime }B^{\prime }}  \label{e50}
\end{equation}%
and%
\begin{equation}
(\square +\frac{1}{2}R)\Psi {}^{ABCD}{}-6\Psi _{MN}{}^{(AB}\Psi
{}^{CD)}{}^{MN}=-\kappa \nabla _{(A^{\prime }}^{(A}\nabla _{B^{\prime
})}^{B}T^{CD)A^{\prime }B^{\prime }}{}.  \label{e51}
\end{equation}

\section{Concluding remarks and outlook}

The curvature quantity $\Lambda $ defined in Ref. [12] always obeys the
relation%
\begin{equation}
\frac{1}{8}R=\chi =3\Lambda ,  \label{cr1}
\end{equation}%
whence, by (\ref{add10}), the equality $\lambda =2\chi $ holds when $T=0$.
In the absence of the cosmological constant, the number of degrees of
freedom of a X-spinor for an empty spacetime is reduced to $10,$ and we
correspondingly have the simplified expansion%
\begin{equation}
\text{X}_{ABCD}=\Psi _{ABCD},  \label{cr2}
\end{equation}%
which still applies to the case of a trace-free $T_{\mu \nu }$ in the
absence of $\lambda .$ Under these circumstances, if the background
spacetime also possesses conformal flatness, then the spinor X$_{ABCD}$ must
be taken as an identically vanishing object. For a conformally flat empty
spacetime in the presence of the cosmological constant, we get from (\ref{X5}%
)%
\begin{equation}
\text{X}_{ABCD}=-\frac{1}{3}\lambda \varepsilon _{A(C}\varepsilon _{D)B},
\label{cr9}
\end{equation}%
which reinstates the relation $\lambda =2\chi $ through%
\begin{equation}
\text{X}_{AB}{}^{AB}=\lambda .  \label{cr12}
\end{equation}

A noteworthy feature of our presentation is the relationship between the
total symmetry of the source $S_{ABCD}{}$ and the conservation law that is
expressed as the covariant divergence freeness of $T_{\mu \nu }.$ We saw
that the contracted derivative (\ref{e42}) comes out at the intermediate
stages of the expansions that yield Eq. (\ref{s3}), but its vanishing has
nevertheless to be thought of as a differential property related to
Einstein's equations as well as to Eqs. (\ref{X1}) and (\ref{e37}). Thus,
one could eventually be led to making the conservation statement upon
keeping track of index configurations. Such results have provided us with a
fresh insight into the situation concerning the derivation of generally
relativistic wave equations as well as with a clear magnification of the
spinor symmetries carried by the geometric structure of general relativity.

The source-free limiting case of Eqs. (\ref{e50}) and (\ref{e51}) is
trivially defined by the statements%
\begin{equation}
(\square +\frac{1}{2}R)\Psi {}_{ABCD}{}-6\Psi _{MN(AB}{}\Psi
{}_{CD)}{}^{MN}=0  \label{cr3}
\end{equation}%
and%
\begin{equation}
(\square +\frac{1}{2}R)\Psi {}^{ABCD}{}-6\Psi _{MN}{}^{(AB}\Psi
{}^{CD)}{}^{MN}=0,  \label{cr5}
\end{equation}%
which control the propagation of gravitons in non-conformally flat vacuum
spacetimes, in addition to coinciding formally with the statements
formulated in Refs. [12,16]. We believe that further insights into the
two-component theory of gravitons could be gained if the formulation just
presented were extended to torsionful spacetime backgrounds.

ACKNOWLEDGEMENTS:

JGC should acknowledge an earlier Editorial Board of Il Nuovo Cimento for
the publication of the ERRATA DOI 10.1393/ncb/i2010-10928-9 concerning the
work of Ref. [31]. CEM would like to acknowledge the Brazilian agency CAPES
for financial support.

Data Availability Statement: the data on the article at issue are entirely
available in the repository ArXiv on the location
https://arxiv.org/abs/2502.12656

\end{document}